# PROGRESS WITH THE UPGRADE OF THE SPS FOR THE HL-LHC ERA

B. Goddard, T. Argyropoulos, H. Bartosik, W. Bartmann, T. Bohl, F. Caspers, K. Cornelis,
H. Damerau, L. Drøsdal, L. Ducimetière, R. Garoby, M. Gourber-Pace, W. Höfle, G. Iadarola,
L. Jensen, V. Kain, R. Losito, M. Meddahi, A. Mereghetti, V. Mertens, Ö. Mete, E. Montesinos,
J. E. Müller, Y. Papaphilippou, G. Rumolo, B. Salvant, E. Shaposhnikova, M. Taborelli, H. Timko,
F. Velotti, CERN, Geneva, Switzerland, E. Gianfelice-Wendt FNAL, Batavia, USA

*Abstract*

The demanding beam performance requirements of the High Luminosity (HL-) LHC project translate into a set of requirements and upgrade paths for the LHC injector complex. In this paper the performance requirements for the SPS and the known limitations are reviewed in the light of the 2012 operational experience. The various SPS upgrades in progress and still under consideration are described, in addition to the machine studies and simulations performed in 2012. The expected machine performance reach is estimated on the basis of the present knowledge, and the remaining decisions that still need to be made concerning upgrade options are detailed.

## INTRODUCTION

The required parameters for HL-LHC have evolved; and at collision in LHC are summarised in Table 1 for 25 ns and 50 ns bunch spacing. It has to be noted that the 50 ns option is highly disfavoured since it produces an unmanageable large number of events per crossing for the LHC experiments. The assumed 10% emittance blow-up and 10% intensity loss between SPS injection and LHC collision [1] define the requirements for the SPS.

Table 1: Main HL-LHC beam parameters at extraction from SPS, for 25 and 50 ns bunch spacing

| Parameter | Unit | 25 ns | 50 ns |
|---|---|---|---|
| Bunch intensity | e11 p+ | 2.4 | 3.9 |
| $\varepsilon_{xy}$ (1 σ rms normalised) | mm.mrad | 2.3 | 2.7 |
| Number of bunches | | 288 | 144 |
| LHC events per crossing | | 169 | 344 |

Limitations in the SPS are longitudinal beam instability, beam loading in the RF systems, Transverse Mode Coupling Instability TMCI, electron cloud, beam losses, kicker element heating through beam coupling impedance and other operational effects including robustness of protection devices, beam instrumentation limitations, electrostatic septum sparking, beam dump outgassing and robustness, and the dynamic range of the halo scraping system. A summary of the present 'accessible' parameter range for the SPS is shown in Fig. 1 with the standard 2012 72/36 (for 25/50 ns) bunches per batch operating points for both 25 and 50 ns, and the high brightness Batch Compression Merging and Splitting (BCMS) beam [2]. Obviously in reality the limitations are not so sharply defined.

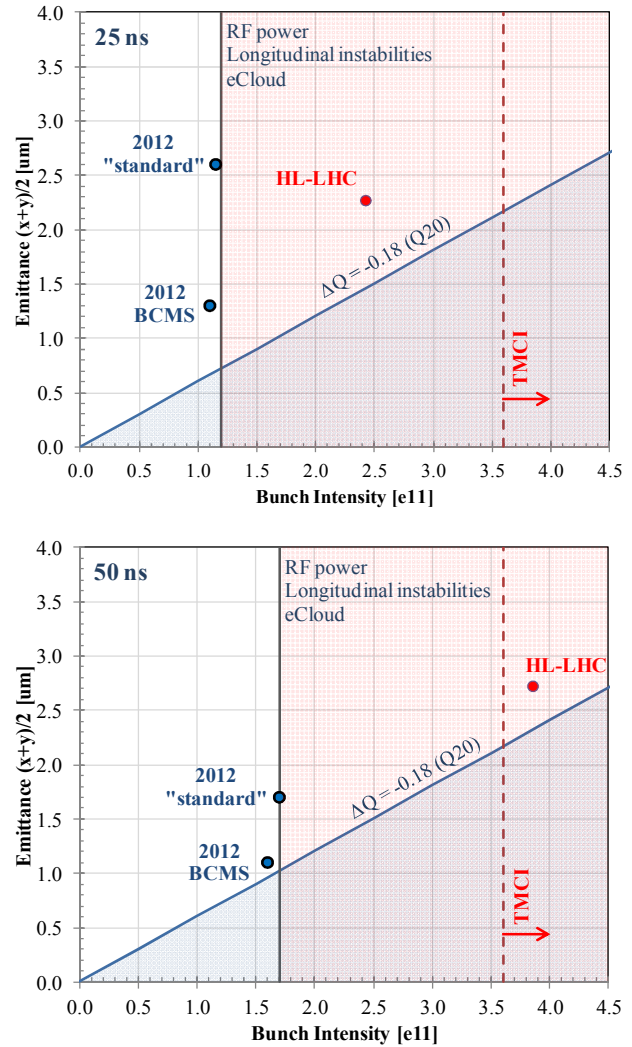

Figure 1: HL-LHC targets, limitations and parameters in SPS for 25 ns (top) and 50 ns (bottom) LHC beams.

## PROGRESS WITH STUDIES IN 2012

In 2012 significant progress was made across the various LIU-SPS study areas.

*Electron cloud*

A dedicated week was used for intensive tests of electron cloud scrubbing and chamber coatings, and to benchmark simulations [3]. The amorphous carbon coating was successfully applied to a chamber inside a dipole and fully validated with beam as a mitigation measure. Promising new scrubbing ideas were tested, including a beam with '20+5' ns spacing [3] produced by injecting a long bunch into two 200 MHz buckets. The intensive simulation effort has allowed an excellent understanding of the various effects and dependencies, for instance the electron density and stripe position as a function of chamber shape, bunch intensity and spacing. In agreement with simulation the electron cloud for 50 ns beams could be fully conditioned away in the MBA dipole type vacuum chambers, Fig. 2, which comprise 50 % of the dipoles.

The sectorisation of the SPS vacuum system is being improved, already locally at sensitive elements like kickers and beam dumps, and in the future in the long arcs to reduce individual pumping section lengths.

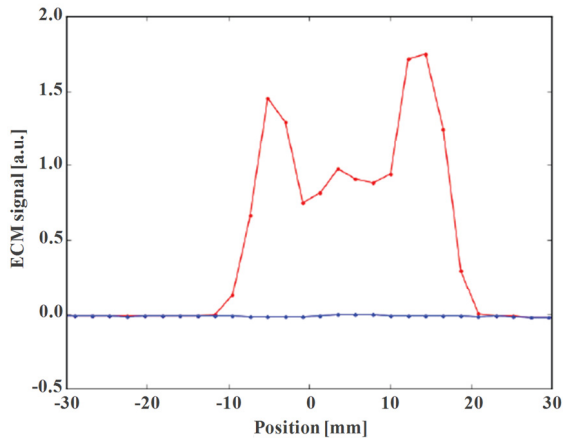

Figure 2: Horizontal distribution of electron flux in strip detector with MBA shaped chamber, measured with 50 ns beam, before (red) and after scrubbing with 25 ns beam.

### Low gamma-transition optics

The new low-gamma transition optics (named Q20) was deployed operationally after intensive testing [4]. The change gave performance improvements in the SPS, with in particular better longitudinal quality of the beam extracted from the SPS and also an increase in transverse brightness from an average of $0.92\times10^{11}$ p+/µm to $1.05\times10^{11}$ p+/µm. Rematching of the extraction systems and transfer lines was required [5].

### Beam transfer efficiency from PS to SPS

Studies on the beam transfer from PS to SPS demonstrated that the beam losses could be halved by optimising the PS bunch rotation in longitudinal phase space. It has been shown experimentally and in simulation for rotated bunches that the bunch length alone cannot be used to predict capture losses, but the longitudinal phase space distribution should be used instead. With higher voltage and optimised timings used for the rotation, the same transmission and bunch length can be maintained for a 40 % larger longitudinal emittance [6], which increases significantly the longitudinal beam stability.

### Longitudinal beam stability

The increase in bunch intensity with 50 ns beam with the normal optics for LHC beams (called Q26) led to longitudinal instabilities at the flat-top prior to extraction to LHC, which frequently slowed down the LHC filling due to the spread in bunch lengths. Switching to Q20 in the middle of 2012 eased the problems, with the onset of longitudinal coupled bunch instability in single RF at twice higher energy than for Q26. For the 25 ns beam, controlled longitudinal emittance blow-up is essential for bunch intensities above $1.2\times10^{11}$ p+. The blow-up is performed by band-limited noise and, with the 800 MHz RF system used in bunch-shortening mode, is very sensitive to the spread in bunch lengths, which gives some systematic effects depending on the bunch lengths coming from the PS [7]. Very careful tuning was required to produce a beam of $1.35\times10^{11}$ p+ per bunch at 25 ns, which is the intensity record for the SPS with the present RF system and Q20 [8].

### Space charge

The availability of the high brightness BCMS beam from the PS and Q20 optics meant that an incoherent space charge tune spread $\Delta Q_y$ of about 0.18 could be achieved without significant emittance growth or losses [8]. This represents an important increase over the previously assumed limit of 0.15, and essentially already reaches the tune spread requirement for HL-LHC, albeit at significantly lower bunch intensity.

## PROGRESS WITH UPGRADES

### Transverse dampers

The new high frequency (intra-bunch) transverse damper demonstrator has been successfully tested in closed-loop mode, damping excited dipole motion [9]; design of a multi-bunch prototype is now under way. This system will potentially fight against electron cloud instabilities, which could be critical e.g, in the case of scrubbing for high intensity.

The existing high power transverse damper system is being upgraded ready for 2014, with new low level controls and dedicated pick-ups being installed.

### RF systems upgrade

To alleviate the limitations posed by beam loading with the 25 ns beam which reduces transmission and increases losses, the upgrades of the different RF systems are continuing. Doubling of the available 800 MHz voltage and the improved low level control are planned to be ready for the SPS restart in 2014. The preparations for the big upgrade of the 200 MHz RF have advanced, with the design and integration of the new building and civil engineering scheduled to be completed end 2014, the prototyping of the power and driver amplifiers under way, the design of the new power couplers in progress and the layout change to the RF cavities and associated SPS straight section defined and launched.

### Machine protection systems

The high beam intensity and brightness mean that the present passive absorbers in the SPS ring and the transfer lines to LHC will no longer survive a full beam impact; neither would they fully protect the downstream accelerator equipment. A series of studies are under way [10] to quantify the expected energy deposition in the different devices, and to design the next generation of absorbers to cope with the new beam parameters.

Additionally, a complete upgrade of the warm magnet interlocking system has started, with replacement of all PLCs and new cabling and sensors.

### Impedance identification and reduction

The final remaining unshielded MKE kicker has been removed and replaced with a serigraphed version which will slightly reduce the overall machine impedance. More importantly, for 2014 this will remove the limitation imposed by the beam heating of this kicker, which was a major issue for efficient scrubbing, as the power deposition is a factor 4 lower in the new kicker. The identification of other impedance sources continues with measurements and increasingly detailed simulations [11].

### Beam instrumentation

A wide range of upgrades and improvements are planned for the SPS beam instrumentation systems.

The entire acquisition electronics of the closed orbit and beam loss monitoring systems will be replaced by a fibre based system by about 2018, which will improve dramatically the acquisition bandwidth and upgrade capacity of both systems. The acquisition system for the fast beam loss monitoring system will be upgraded for maintenance and performance reasons.

Transverse profile measurements will be improved with a new generation of precise fast wire-scanners; a prototype has been designed and is being constructed for beam testing during 2014. The existing rest-gas ionisation monitors are being upgraded allowing for multiple profile measurements through the SPS cycle. The synchrotron radiation telescope system is being improved to allow transverse beam size measurements at high energy.

Bunch-by-bunch tune measurement will be deployed in LS1, and investigations into an improved head-tail monitor acquisition system to cover the short bunch-length at extraction and provide both transverse planes simultaneously.

Finally, the analogue bandwidth of the present fast beam current transformers will be improved to increase the measurement accuracy, crucial for studying instabilities of individual LHC bunches.

### Project baseline reviews

A series of project reviews was organised in early 2013 to examine remaining open questions and also to limit the upgrade possibilities to a much more focused baseline.

The existing scraper system will be retained, with a magnetic bump system design as potential backup [12].

The existing extraction kicker system will also remain, with some improvements to cooling and a reduction in the total number of kickers, but no further new kicker developments.

The internal beam dump absorber block will be upgraded to withstand full LIU beam [13], and an external block in an existing transfer line investigated for dumping of high energy beams.

The pumping of the ZS electrostatic septa will be improved and remote voltage modulation implemented.

No general closed orbit correction system for high energy is required, but local correction of the extraction orbit will be investigated using the extraction bumpers, and transfer lines stability studies will continue [14].

## CONCLUSION

The upgrade of the SPS for the HL-LHC era is well under way, with most uncertainties on the project baseline resolved. 2012 has been a very important year in terms of simulations, beam studies and results, with the new Q20 optics now operational, extensive tests made for electron cloud and an improved understanding of the machine limitations.

The major remaining decision for the project concerns the choice between scrubbing and amorphous carbon chamber coating as mitigation against electron cloud build-up. This will need tests in 2014 to answer, since the main unknown for scrubbing is the length of time needed to recover full performance after a long (18 month) technical stop which will involve extensive opening of the vacuum system.

Alongside electron cloud, the longitudinal beam stability is a critical performance aspect which will benefit from the major RF system upgrades under way, and also from the continuing efforts to reduce impedance.

If all the planned upgrades have the expected beneficial effects, and if no other unexpected limitations manifest themselves at higher intensities, the SPS should be in a good position to satisfy the HL-LHC requirements over the next two decades.